\documentclass[aip,jcp,reprint,amsmath,amssymb,floatfix,nofootinbib]{revtex4-1}
\usepackage{amsmath}
\usepackage[utf8]{inputenc}
\usepackage[version=4]{mhchem}
\usepackage{graphicx}
\usepackage{tikz}
\usepackage{fix-cm}
\usepackage{color}
\usepackage[normalem]{ulem}
\usepackage{natbib}
\usepackage{epstopdf}

\newcommand{\transp}{\mathrm{T}}

\renewcommand{\phi}{\varphi}
\renewcommand{\theta}{\vartheta}

\renewcommand{\b}{\mathbf}

\newcommand{\ket}[1]{\vert #1 \rangle}

\newcommand{\Dbraket}[2]{\langle #1 \hspace{.10em} \vert \hspace{.10em}  #2 \rangle}
\newcommand{\Tbraket}[3]{\langle #1 \hspace{.10em} \vert \hspace{.10em} #2 \hspace{.10em} \vert \hspace{.10em} #3 \rangle}

\newcommand{\ident}{\mathbb{I}}

\newcommand{\real}{\mathbb{R}} 

\newcommand{\rf}{\mathrm{R}}

\newcommand{\J}{\b{A}}

\newcommand{\nucl}{\b{R}}

\newcommand{\bsym}{\boldsymbol} 

\begin{document}
\title{Crossing conditions in coupled cluster theory}
\author{Eirik F.~Kj{\o}nstad}
\affiliation{Department of Chemistry, Norwegian University of Science and Technology, 7491 Trondheim, Norway} 
\affiliation{Department of Chemistry and the PULSE Institute, Stanford University, Stanford, California 94305, USA}
\author{Rolf H.~Myhre}
\affiliation{Department of Chemistry, Norwegian University of Science and Technology, 7491 Trondheim, Norway} 
\affiliation{Department of Chemistry and the PULSE Institute, Stanford University, Stanford, California 94305, USA}
\author{ Todd J.~Martínez}
\affiliation{Department of Chemistry and the PULSE Institute, Stanford University, Stanford, California 94305, USA}
\author{Henrik Koch}
\email[Corresponding author: ]{henrik.koch@ntnu.no}
\affiliation{Department of Chemistry, Norwegian University of Science and Technology, 7491 Trondheim, Norway} 
\affiliation{Department of Chemistry and the PULSE Institute, Stanford University, Stanford, California 94305, USA}

\setlength{\arraycolsep}{3pt}

\begin{abstract}
We derive the crossing conditions at conical intersections between electronic states in coupled cluster theory, and show that if the coupled cluster Jacobian matrix is nondefective, two (three) independent conditions are correctly placed on the nuclear degrees of freedom for an inherently real (complex) Hamiltonian. Calculations using coupled cluster theory on an $2 \,^{1}A' / 3 \,^{1}A'$ conical intersection in hypofluorous acid illustrate the nonphysical artifacts associated with defects at accidental same-symmetry intersections. In particular, the observed intersection seam is folded about a space of the correct dimensionality, indicating that minor modifications to the theory are required for it to provide a correct description of conical intersections in general. We find that an accidental symmetry allowed $1 \; {^{1}}A{''} / 2 \; {^{1}}A{''}$ intersection in hydrogen sulfide is properly described, showing no artifacts as well as linearity of the energy gap to first order in the branching plane. 
\end{abstract}

\maketitle

\section{Introduction} 
A realistic description of nuclear motion in excited electronic states requires reliable predictions of the energies of such states and the nonadiabatic coupling between them. Electronic state degeneracies, more commonly referred to as conical intersections, are now widely recognized to play a prominent role in such dynamics, for instance in photochemistry.\citep{Zhu2016} The successful use of \emph{ab initio} quantum chemical methods to predict dynamics is challenging, however. Sufficient accuracy often coincides with prohibitive computational cost, and the multireference problem hinders the global accuracy of practical theories.\citep{Lyakh2012} It has nevertheless been shown that dynamics simulations involving conical intersections, for both isolated and condensed-phase systems, can be predictive and offer novel insights into the mechanisms following photoexcitation.\citep{Ben2000,Burghardt2004,Toniolo2004,Levine2007,Wolf2017}

In the early days of quantum mechanics, von Neumann and Wigner derived the conditions necessary for two electronic states to become degenerate.\citep{vonNeumann1929} They realized that two conditions are satisfied at such an intersection, in the sense that $u(\nucl) = 0$ and $v(\nucl) = 0$, where $u$ and $v$ are independent functions of the internal nuclear coordinates $\nucl$.\citep{Yarkony1996} Although the proof and its interpretation was once a subject of some controversy,\citep{Naqvi1972} there now exist many independent mathematical proofs of their original insight.\citep{Teller1937,LonguetHiggins1975,Mead1979} The number of conditions has implications for the structure of conical intersections. Two conditions are expected to be satisfied in a subspace of dimension $N-2$, where $N$ is the number of internal nuclear degrees of freedom.\citep{Anton2010} In this subspace, known as the intersection seam, the degeneracy is preserved; in its complement, the branching plane, the surfaces adopt the shape of two facing cones (giving the intersections their name, conical).\citep{Teller1937} Note that the number of conditions is not always two. When effects that render the Hamiltonian inherently complex are accounted for, the two conditions become three, for instance.\citep{vonNeumann1929} 

For diatomics, the proof implies the noncrossing rule, which states that  states of the same symmetry cannot intersect. This is because the likelihood that two conditions are satisfied by varying one parameter, the distance between the atoms, is vanishingly small, and it will therefore never happen in practice.\citep{Mead1979} 

The conical intersections of approximate theories may be qualitatively incorrect if they do not faithfully reproduce the crossing conditions. Considering the eigenvalue equation associated with a nonsymmetric matrix, Hättig\citep{Hattig2005} noted that it seems to enforce three conditions, rather than two, when two eigenvalues become equal. This would imply that matrix symmetry is needed to obtain the correct number of conditions, thereby ruling out coupled cluster response theory\citep{Koch1990b,Stanton1993a} as a viable model at conical intersections. Soon afterwards, Köhn and Tajti\citep{Kohn2007} found complex energies and parallel eigenvectors using coupled cluster theory, truncated after singles and doubles (CCSD)\citep{Purvis1982} and triples (CCSDT)\citep{Noga1987,Noga1988} excitations, at an intersection in formaldehyde. More recently, complex energies were encountered in nuclear dynamics simulations using the perturbative doubles coupled cluster (CC2)\citep{Christiansen1995} model.\citep{Plasser2014} 

\begin{figure*}
  \centering
 \includegraphics[width=0.7\linewidth]{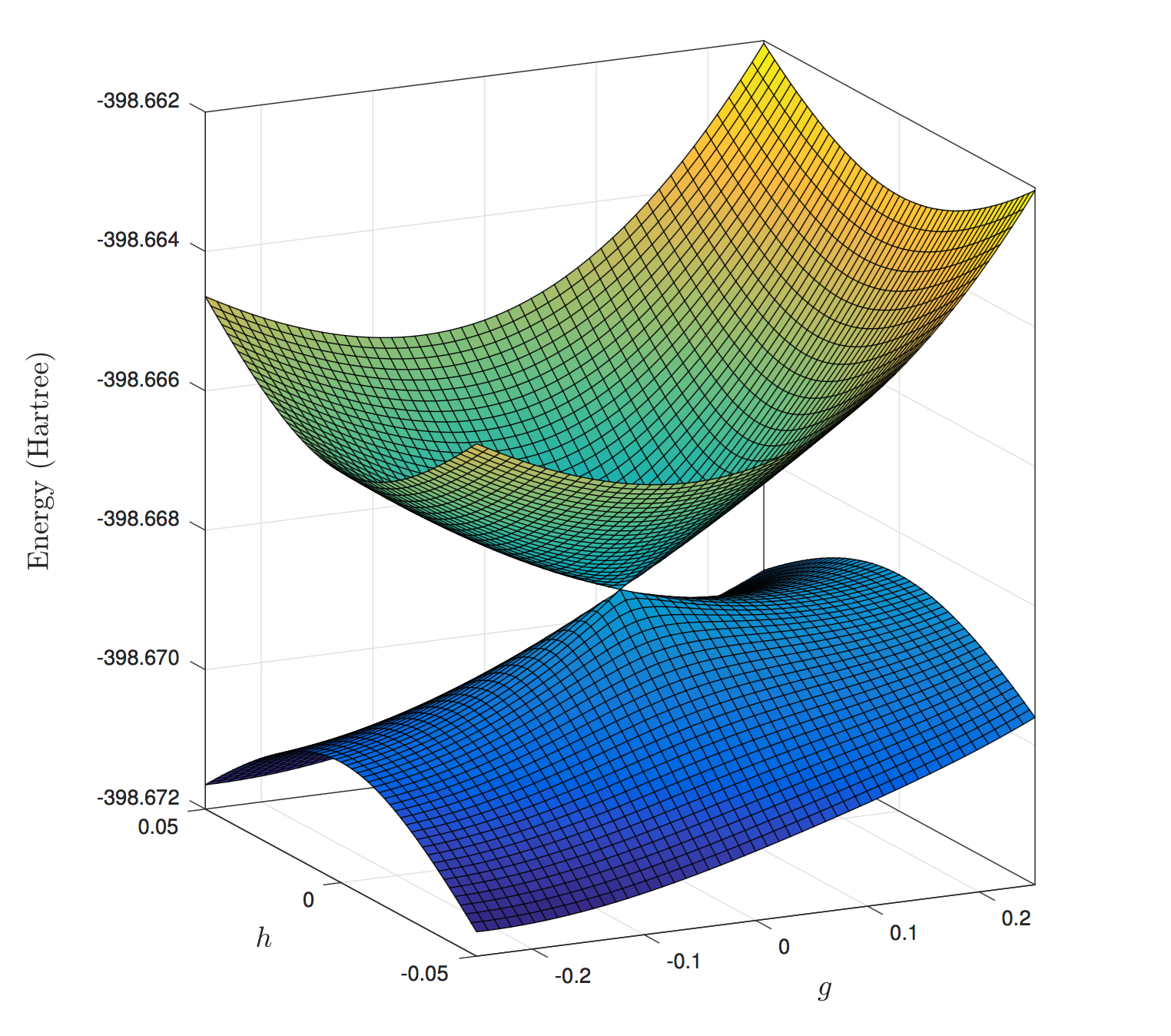}
  \caption{An accidental symmetry allowed conical intersection of the $1 \; {^{1}} A{''}$ and $2 \; {^{1}} A{''}$ states of \ce{SH2} using CCSD/aug-cc-pVDZ.\label{fig:symmetryallowed}}
\end{figure*}

In the present contribution, the crossing conditions of nonsymmetric matrices are reconsidered, with particular attention given to the case of coupled cluster theory. We show that nonsymmetric matrices reproduce the crossing conditions of quantum mechanics, provided the matrices are nondefective. In addition, we  argue that it is misleading to identify the intersection of the method with the $(N-3)$-dimensional space resulting from Hättig's conditions, as some authors have.\citep{Hattig2005,Kohn2007,Gozem2014} In light of coupled cluster theory's behavior in the limit where all excitation operators enter the cluster operator, the conical intersection is more appropriately identified with an $(N-1)$-dimensional space, also discussed by Hättig,\citep{Hattig2005} that is folded about a space of the correct dimensionality. 

In the current contribution, we restrict our attention to intersections between excited states and defer  intersections between the ground and excited states to a later publication. The observations in this paper form the basis for a modified coupled cluster model that is nondefective and therefore able to describe conical intersections between excited states of the same symmetry.\citep{Kjoenstad2016} 

\section{Coupled cluster crossing conditions} \label{sec:crossing}
Let us first derive the crossing conditions in quantum mechanics, where we follow closely the argument given by Teller.\citep{Teller1937} Denote the two electronic states of interest by $\Psi_1$ and $\Psi_2$, where
\begin{align}
H \, \Psi_k = E_k \, \Psi_k, \quad k = 0,1,2,\ldots
\end{align}
Here and in the subsequent discussions, $H$ is the clamped nuclei Hamiltonian expressed in the determinantal basis. Let us define a reduced space representation of $H$ in an orthonormal basis that spans $\Psi_1$ and $\Psi_2$, say $\Phi_1$ and $\Phi_2$:
\begin{align}
\b{H} \, \b{r}_k = E_k \, \b{r}_k, \quad H_{ij} = \Dbraket{\Phi_i}{H \, \Phi_j}, \quad i,j = 1,2.
\end{align}
The eigenvalues of $\b{H}$ then satisfy
\begin{align}
E_2 - E_1 = \sqrt{(H_{22} - H_{11})^2 + 4 \, H_{12}^2}. \label{eq:eigenvalues}
\end{align}
As long as the basis functions $\Phi_1$ and $\Phi_2$ are real, $\b{H}$ will be symmetric ($\b{H} = \b{H}^T$) because its elements will be real for all nuclear coordinates $\nucl$.
Since $E_2 - E_1$ vanishes by definition at an intersection, Eq.~\eqref{eq:eigenvalues} gives us the crossing conditions
\begin{align}
\begin{split}
0 &= u(\b{R}) = H_{22}(\b{R}) - H_{11}(\b{R}), \\
0 &= v(\b{R}) = H_{12}(\b{R}). 
\end{split} \label{eq:crossing}
\end{align}
If $u(\nucl)$ and $v(\nucl)$ are expanded to first order about an intersection point, $\nucl_0$, we expect to obtain solutions to Eq.~\eqref{eq:crossing}, $\nucl$, belonging to a subspace of dimension $N-2$, where $N$ is the number of nuclear degrees of freedom.\citep{Anton2010}

An analogous proof can be attempted in the framework of coupled cluster theory. The excitation energies $\omega_k$ in this model are the eigenvalues of the nonsymmetric coupled cluster Jacobian matrix $\J$:
\begin{align}
\J \, \b{r}_k = \omega_k \, \b{r}_k, \quad k = 0,1,2,\ldots.
\end{align}
In terms of the cluster operator ${T = \sum_{\mu > 0} t_\mu \tau_\mu}$, a sum of excitation operators $\tau_\mu$ weighted by amplitudes $t_\mu$ obtained from the amplitude equations,
\begin{align}
A_{\mu\nu} = \Tbraket{\mu}{e^{-T} (H-E_0) \, e^T}{\nu}, \quad \ket{\nu} = \tau_\nu \, \ket{\rf},
\end{align}
where $E_0 = \Tbraket{\rf}{e^{-T} H e^T}{\rf}$, $\ket{\rf}$ is the Hartree-Fock determinant, $\tau_0 = \ident$, and $\mu,\nu \geq 0$. Although $\J$ is usually defined only for $\mu, \nu > 0$,\citep{Koch1990,Stanton1993a} we include the reference terms here. This is useful because $\J$ is then directly related to $H$ in the limit of a complete cluster operator.

We denote the left eigenvectors of $\J$ by $\b{l}_k$, and define a reduced representation of $\J$, given in a biorthonormal basis $\{ \bsym{\lambda}_i, \bsym{\rho}_i \}_{i = 1,2}$ of the space spanned by $\{ \b{l}_i, \b{r}_i \}_{i = 1,2}$:
\begin{align}
\b{J} = \begin{pmatrix} 
J_{11} & J_{12} \\
J_{21} & J_{22}
\end{pmatrix}, \quad J_{ij} = \bsym{\lambda}_i^\transp \, \J \, \bsym{\rho}_j, \quad \bsym{\lambda}_i^\transp\bsym{\rho}_j = \delta_{ij}. \label{eq:redJacDef}
\end{align}
The eigenvalues of $\b{J}$ satisfy
\begin{align}
\omega_2 - \omega_1 = \sqrt{(J_{22} - J_{11})^2 + 4 \, J_{12}\,J_{21}}. \label{eq:jacSecular}
\end{align}

At this point in the proof, we see that the crossing conditions cannot be inferred from Eq.~\eqref{eq:jacSecular}, because $J_{12}\,J_{21}$ may be negative for nonsymmetric $\J$. To proceed, we have to consider the linear independence of the eigenvectors of $\J$. Note that this complication is not present in symmetric theories, where the eigenvectors are linearly independent due to their orthonormality.

The Hamiltonian is Hermitian, and its eigenvalues are consequently real, its eigenvectors orthogonal. It always has a diagonal representation, and its eigenvectors always span the entire Hilbert space.\citep{Teschl2014} Orthogonality is lost for nonsymmetric matrices (such as $\J$), and there is no guarantee that its eigenvectors span the entire space. If they do, $\J$ is called nondefective and can be written
\begin{align}
\J = \sum_k \omega_k \, \b{r}_k \, \b{l}_k^\mathrm{T}, \quad \b{I} = \sum_k \b{r}_k \, \b{l}_k^\mathrm{T}. \label{eq:jacSpectral}
\end{align}
Equivalently, $\J$ is nondefective if it can be diagonalized: that is, if there exists an $\b{M}$ such that $\J = \b{M}^{-1} \bsym{\omega} \, \b{M}$, where $\bsym{\omega}$ is a diagonal matrix. The eigenvalue associated with a matrix defect is  known as a defective eigenvalue.\citep{Golub2012}

Symmetric matrices are nondefective because they can always be diagonalized. Nonsymmetric matrices, on the other hand, are guaranteed to be nondefective only at $\nucl$ where the eigenvalues are distinct. When the eigenvalues are distinct, the associated eigenvectors can be shown to be linearly independent.\citep{Anton2010} Nonsymmetric matrices may therefore become defective at intersections, though this is not neccessarily the case. For instance, coupled cluster theory is nondefective for a complete cluster operator, at which point $\J$ is a matrix representation of $H$.\citep{Koch1990}

The crossing conditions may be derived by assuming that $\b{J}$ is nondefective. Denoting the right and left eigenvectors of $\b{J}$ by $\b{q}_k$ and $\b{p}_k$, for $k = 1,2$, we find that
\begin{align}
\b{J} = \sum_{k = 1,2} \omega \, \b{q}_k \, \b{p}_k^\transp = \omega \sum_{k = 1,2} \, \b{q}_k \, \b{p}_k^\transp = \omega \, \b{I}, \label{eq:redJacDiagonal}
\end{align}
from which it follows, by Eqs.~\eqref{eq:redJacDef} and \eqref{eq:redJacDiagonal}, that
\begin{align}
\begin{split}
0 &= u(\b{R}) = J_{22}(\b{R}) - J_{11}(\b{R}), \\
0 &= v(\b{R}) = J_{12}(\b{R}), \\
0 &= w(\b{R}) = J_{21}(\b{R}). \label{eq:ccCrossing}
\end{split}
\end{align}
These crossing conditions were first postulated by Hättig, representing what he named a true intersection.\citep{Hattig2005} The number of conditions has led some to erronously conclude that the branching plane for a two-state intersection is always three-dimensional in nonsymmetric theories.\citep{Hattig2005,Kohn2007}

\begin{figure*}
\includegraphics[width=\linewidth]{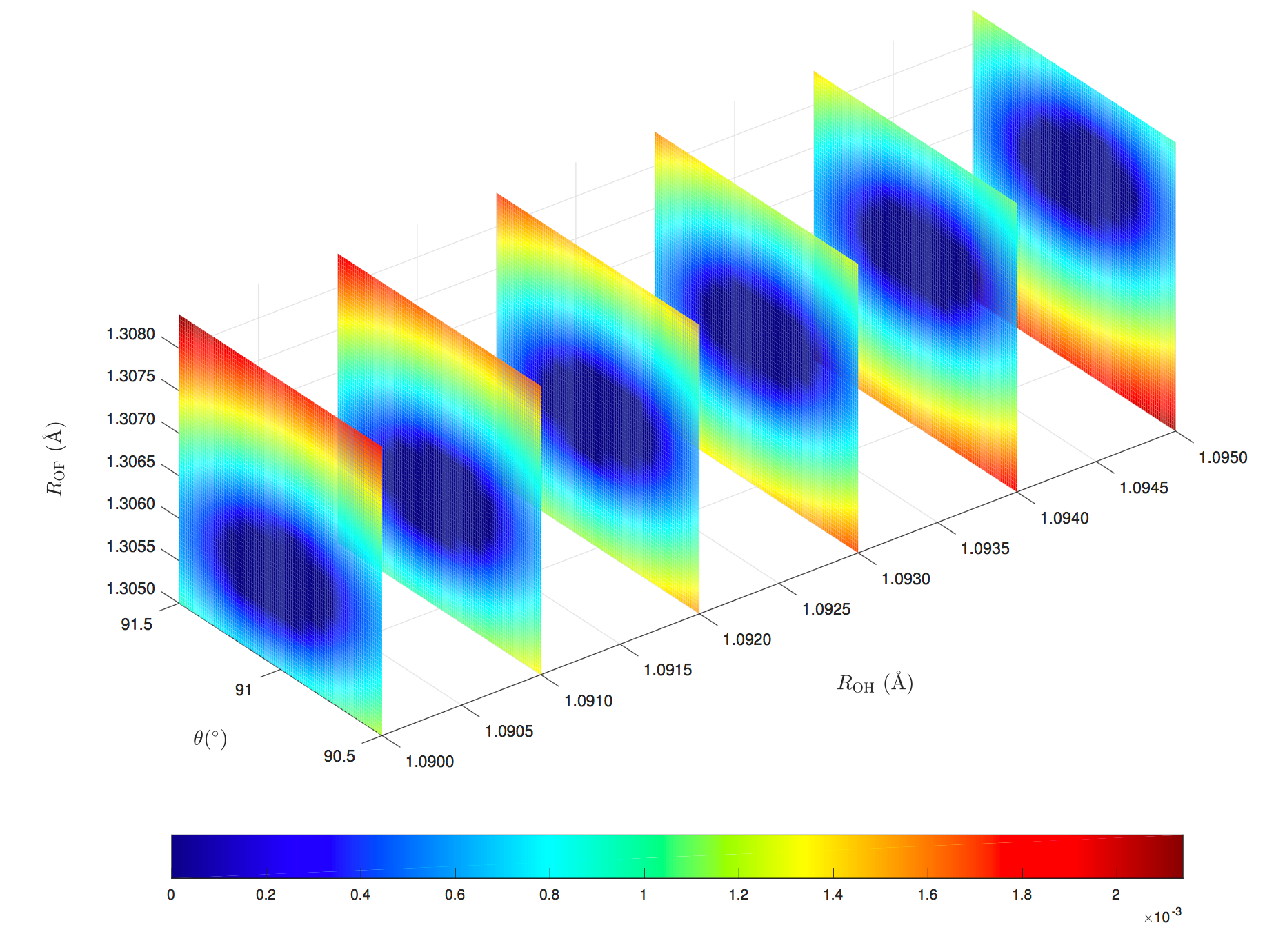}
\caption{\label{fig:seam_hof} The 2$^1$A$^\prime$/3$^1$A$^\prime$ LR-CCSD/aug-cc-pVDZ intersection seam in \ce{HOF}. The coloring of each plot gives the real part of the energy difference, in hartrees, between the 2$^1$A$^\prime$ and 3$^1$A$^\prime$ electronic states.} \label{fig:realPartHOF}
\end{figure*}

\section{Intersection dimensionality} \label{sec:dimensionality}
Let us show that a nondefective $\b{J}$ implies the correct intersection seam dimensionality. Suppose that $\b{J}$ is nondefective in some region $\mathcal{S} \subseteq \real^N$, where, for illustrative purposes, we let $\nucl$ space be three-dimensional ($N = 3$), as is true for triatomic systems. Then, if $u$, $v$, and $w$ are independent functions of $\nucl$, each condition defines a plane, say $\mathcal{A}$, $\mathcal{B}$, and $\mathcal{C}$, which might be expected to intersect at a point. This is not what occurs, however. The situation is instead one where two of the planes intersect to form a curve ($\mathcal{A} \cap \mathcal{B}$) in the third plane ($\mathcal{C}$). To show this, we let $N$ be general and consider the two sets
\begin{align}
\mathcal{J} &= \{ \b{R} \in \mathcal{S} : u(\b{R}) = 0, v(\b{R}) = 0 \}, \\
\mathcal{I} &= \{ \b{R} \in \mathcal{S} : u(\b{R}) = 0, v(\b{R}) = 0, w(\b{R}) = 0 \}.
\end{align}
We wish to prove that $\mathcal{J} = \mathcal{I}$. Clearly, $\mathcal{I} \subseteq \mathcal{J}$, so it is sufficient to show that $\mathcal{J} \subseteq \mathcal{I}$. Suppose on the contrary that $\mathcal{J} \nsubseteq \mathcal{I}$; that is, suppose there is an $\nucl$ in $\mathcal{J}$ that is not in $\mathcal{I}$. Then $\b{J}$ can be written
\begin{align}
\b{J} = \begin{pmatrix} J_{11} & 0 \\
J_{21} & J_{11} \end{pmatrix}. \label{eq:contradiction}
\end{align}
This is a defective matrix: it has one eigenvector, $(0\; 1)^T$, associated with the doubly degenerate eigenvalue $J_{11}$. In other words, $\mathcal{J} \nsubseteq \mathcal{I}$ leads to a contradiction (that $\b{J}$ is defective at $\nucl$), so $\mathcal{J} \subseteq \mathcal{I}$ and hence $\mathcal{J} = \mathcal{I}$. We have thus shown that two independent conditions are enforced at conical intersections, provided $\J$ is nondefective in the subspace of the intersecting eigenvectors. 

\begin{figure*}
\includegraphics[width=\linewidth]{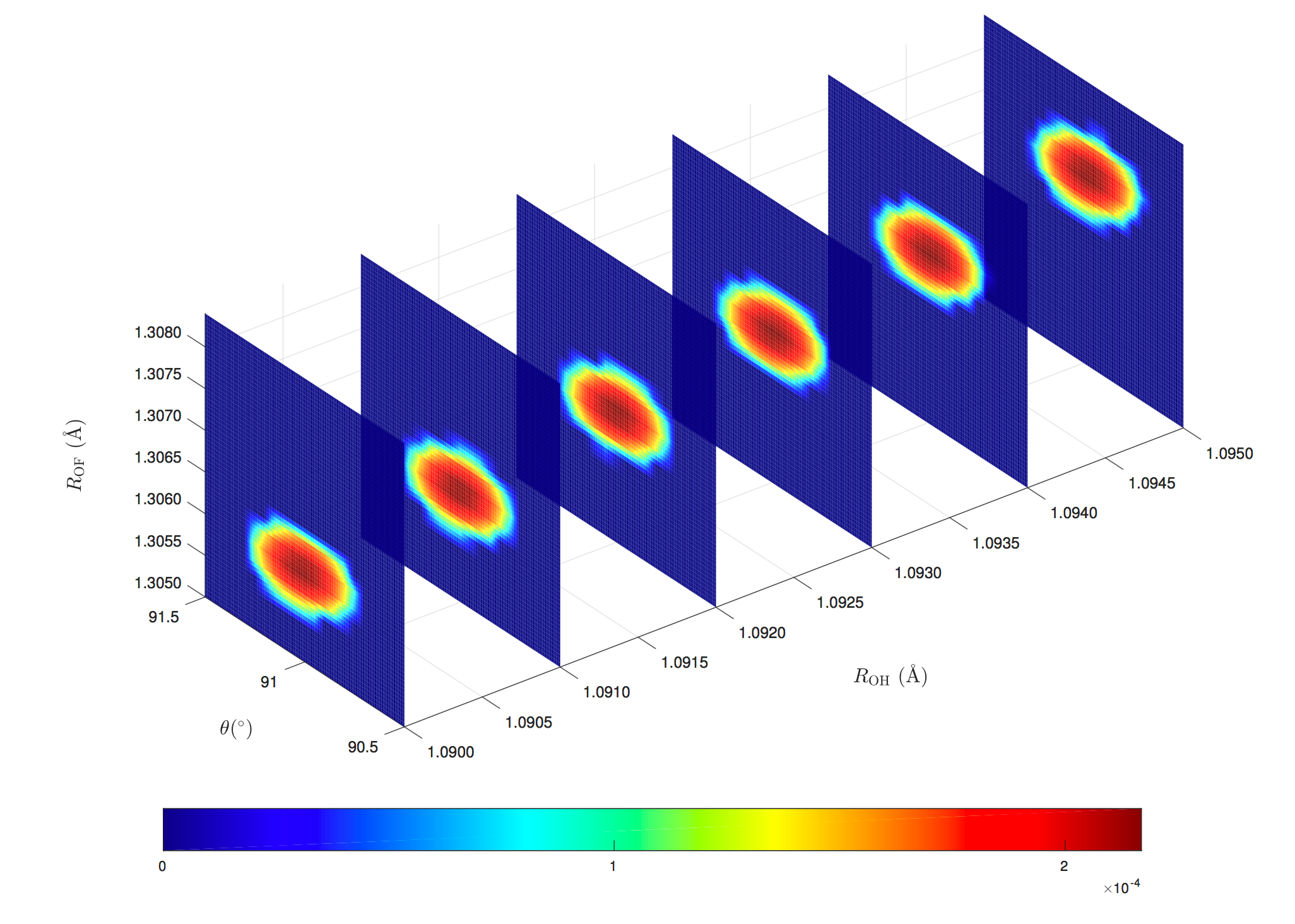}
\caption{As in Figure \ref{fig:realPartHOF}, but showing the imaginary part of the energy difference between the $2 \; {^{1}}A'$ and $3 \; {^{1}}A'$ electronic states.}
\label{fig:imagPartHOF} 
\end{figure*}

That the number of independent conditions is two can also be understood by noting that $\b{J}$ is similar to a symmetric matrix in $\mathcal{S}$. To be nondefective the matrix must be diagonalizable, $\b{J} = \b{M}^{-1} \bsym{\omega} \, \b{M}$. This observation leads to an alternative but equivalent proof. The columns of $\b{M}$ are the right eigenvectors, $\b{x}$ and $\b{y}$, of the matrix $\b{J}$: 
\begin{align}
\b{M} = \begin{pmatrix} x_1 & y_1 \\ x_2 & y_2 \end{pmatrix}, \quad \det \b{M} = x_1 \, y_2 - y_1 \, x_2.
\end{align}
In terms of $\b{M}$ and $\bsym{\omega}$, the crossing conditions read
\begin{align}
u(\nucl) &=  \frac{2 \, \Delta \omega}{\det \b{M}}(x_1 \, y_2 + y_1 \, x_2), \\
v(\nucl) &= \frac{2 \, \Delta \omega}{\det \b{M}}(- y_1 \, y_2) , \\
w(\nucl) &= \frac{2 \, \Delta \omega}{\det \b{M}}(x_1 \, x_2) . \label{eq:w_explicit}
\end{align}
Now suppose it were true that $w(\nucl) \neq 0$ while ${u(\nucl) = 0}$ and ${v(\nucl) = 0}$. From Eq.~\eqref{eq:w_explicit} we then see that $\Delta \omega \neq 0$. It follows that ${y_1 \, y_2 = 0}$ and ${x_1 \, y_2 = - y_1 \, x_2}$. If $y_1 = 0$, then ${\det \b{M} = 0}$; if $y_1 \neq 0$, then $y_2 = 0$, implying ${\det \b{M} = 0}$. This contradicts the fact that the determinant of $\b{M}$ is nonzero because $\b{J}$ is nondefective. We thus conclude that ${u(\nucl) = 0}$ and $v(\nucl) = 0$ together imply that ${w(\nucl) = 0}$.

In quantum mechanics, three conditions are satisfied at an intersection when $\b{H}$ is inherently complex. This is because the off-diagonal element $H_{12}$ cannot be assumed real, giving the modified crossing conditions\citep{Matsika2001}
\begin{align}
\begin{split}
0 &= u(\b{R}) = H_{22}(\b{R}) - H_{11}(\b{R}), \\
0 &= v_1(\b{R}) = \mathrm{Re} \, H_{12}(\b{R}), \\
0 &= v_2(\b{R}) = \mathrm{Im} \, H_{12}(\b{R}).
\end{split}
\end{align}
Retracing the steps made for real $\J$, we find the coupled cluster crossings conditions for complex $\J$ to be
\begin{align}
\begin{split}
0 &= u_1(\b{R}) = \mathrm{Re} \, (J_{22}(\b{R}) - J_{11}(\b{R})), \\
0 &= u_2(\b{R}) = \mathrm{Im} \, (J_{22}(\b{R}) - J_{11}(\b{R})), \\
0 &= v_1(\b{R}) = \mathrm{Re} \, J_{12}(\b{R}), \\
0 &= v_2(\b{R}) = \mathrm{Im} \, J_{12}(\b{R}), \\
0 &= w_1(\b{R}) = \mathrm{Re} \, J_{21}(\b{R}), \\
0 &= w_2(\b{R}) = \mathrm{Im} \, J_{21}(\b{R}).
\end{split}
\end{align}
We assume that the eigenvalues $\omega_i$ are real and nondefective. If we then let
\begin{align}
\mathcal{J} &= \{ \b{R} \in \mathcal{S} : u_1(\b{R}) = 0, v_i(\b{R}) = 0 \}, \\
\mathcal{I} &= \{ \b{R} \in \mathcal{S} : u_i(\b{R}) = 0, v_i(\b{R}) = 0, w_i(\b{R}) = 0 \},
\end{align} 
where $i = 1,2$, $\mathcal{J} = \mathcal{I}$ can be shown as follows. Clearly, $\mathcal{I}\subseteq \mathcal{J}$. To prove that $\mathcal{J} \subseteq \mathcal{I}$, we suppose that there is an $\nucl \in \mathcal{J}$ that is not in $\mathcal{I}$. Then we find
\begin{align}
\Delta \omega = i \; \lvert \, \mathrm{Im} \, (J_{11}-J_{22}) \, \rvert
\end{align}
from Eq.~\eqref{eq:jacSecular}.
As the $\omega_i$ are real, $\mathrm{Im} \, (J_{11}-J_{22}) = 0$. But then
\begin{align}
\b{J} = \begin{pmatrix} J_{11} & 0 \\ J_{21} & J_{11} \end{pmatrix},
\end{align}
contradicting the assumption that $\b{J}$ is nondefective. For complex $\b{H}$, three independent conditions are thus enforced at conical intersections, provided $\J$ is nondefective in the subspace of the intersecting eigenvectors. 

The reduced number of independent conditions explain some facts not accounted for in earlier analyses. One is that $\J$ is nonsymmetric even in the full configuration interaction limit, meaning that the three conditions in Eq.~\eqref{eq:ccCrossing} are satisfied at its intersections. Yet the eigenvalues of $\J$ and $H$, and therefore also their conical intersections, are the same in this limit.\citep{Koch1990} All of this stems from the fact that symmetric theories can be disguised as nonsymmetric. A symmetric matrix, say $\b{B}$, can be made nonsymmetric by a similarity transform by a nonunitary matrix, say $\b{C}$:
\begin{align}
\b{B} \mapsto \b{B}' = \b{C}^{-1} \, \b{B} \, \b{C}.
\end{align}
But the intersections of $\b{B}$ and $\b{B}'$ are identical, because such a transformation does not change the eigenvalues.\citep{Anton2010} There are only two independent conditions in both cases, as $\J$ is similar to $H$ in the limit, and $\b{B}'$ similar to $\b{B}$. Both are nondefective for all $\nucl$. 

Note that similarity of $\J$ to a symmetric matrix, which would amount to a complete symmetrization of the theory, is too strict a criterion. To correctly predict intersections, only the representation of $\J$ in the space spanned by the two eigenvectors, that is, $\b{J}$, needs to be nondefective.

\section{Energy gap linearity in the branching plane: a perturbation theoretical analysis} \label{sec:perturbation}
In the present section we adapt the analysis of Zhu and Yarkony for the special case of coupled cluster theory,\citep{Zhu2016} and examine the behavior of the energy gap in the branching plane close to the conical intersection. 

We let $\nucl_0$ be a point of intersection and expand $\J$ at $\nucl = \nucl_0 + \delta \nucl$:
\begin{align}
\begin{split}
\J(\nucl) &= \J(\nucl_0) + \sum_{\alpha=1}^{3 N_\mathrm{nuc}} \delta R_\alpha \frac{\partial \J(\nucl)}{\partial R_\alpha} \Big\vert_{\nucl_0} \\
&+ \frac{1}{2} \sum_{\alpha, \beta = 1}^{3 N_\mathrm{nuc}} \delta R_\alpha \frac{\partial^2 \J(\nucl)}{\partial R_\alpha \partial R_\beta} \Big\vert_{\nucl_0} \delta R_\beta + \ldots
\end{split}
\end{align}
Then we define a fixed matrix $\b{M}$, whose columns are the right eigenvectors of $\J$ at $\nucl_0$. Let us partition $\J$ into a block of intersecting states ($I$) and its complement ($C$), and transform it to the eigenvector basis at $\nucl_0$:
\begin{align}
\b{M}^{-1} \J \,  \b{M} = \begin{pmatrix} \J^{I,I} & \J^{I,C} \\ \J^{C,I} & \J^{C,C} \end{pmatrix}.
\end{align}
We assume that $\J$ is nondefective in a neighborhood of $\nucl_0$, implying in particular that $\b{M}$ is invertible.
Folding the $C$ block into the $I$ block, the eigenvalue problem for the intersecting states becomes
\begin{align}
(\J^{I,I} + \J^{I,C} (\omega_k - \J^{C,C})^{-1} \J^{C,I} - \omega_k) \, \b{r}_k^I = 0. \label{eq:folded}
\end{align}
As $\b{M}$ consists of the eigenvectors of $\J$ at $\nucl_0$, we have
\begin{align}
A^{I,I}_{ij} &= \delta_{ij} \, \omega_i^{\nucl_0} + \sum_\alpha \delta R_\alpha \frac{\partial A^{I,I}_{ij}}{\partial R_\alpha} \Big\vert_{\nucl_0} + \ldots \\
A^{I,C}_{ij} &= \sum_\alpha \delta R_\alpha \frac{\partial A^{I,C}_{ij}}{\partial R_\alpha} \Big\vert_{\nucl_0} + \ldots \\
A^{C,I}_{ij} &= \sum_\alpha \delta R_\alpha \frac{\partial A^{C,I}_{ij}}{\partial R_\alpha} \Big\vert_{\nucl_0} + \ldots \\
A^{C,C}_{ij} &= \delta_{ij} \, \omega_i^{\nucl_0} + \sum_\alpha \delta R_\alpha \frac{\partial A^{C,C}_{ij}}{\partial R_\alpha} \Big\vert_{\nucl_0} + \ldots
\end{align}
Both $\J^{I,C}$ and $\J^{C,I}$ are first order in $\delta R_\alpha$. The second term in Eq.~\eqref{eq:folded} has no contributions to first order, and the equation therefore reads, to first order,
\begin{align}
\sum_j \Bigl(\delta_{ij} (\omega_i^{\nucl_0}-\omega_i) + \sum_\alpha \delta R_\alpha \frac{\partial A^{I,I}_{ij}}{\partial R_\alpha} \Big\vert_{\nucl_0} \Bigr) r_{i,j}^I = 0, \label{eq:first_order_eig}
\end{align} 
 where $r_{i,j}^I$ is the $j$th element of $\b{r}_i^I$. Denote the degenerate eigenvalue at $\b{R}_0$, $\omega^{\nucl_0}$, by $\omega$. Restricting ourselves to the $I$ block, $i = 1,2$,  we can write Eq.~\eqref{eq:first_order_eig} as
\begin{align}
\begin{pmatrix} \omega+(\b{s} - \b{g})\cdot \delta \b{R} & \b{h}^{1,2}\cdot \delta \b{R} \\ \b{h}^{2,1}\cdot \delta \b{R} & \omega+ (\b{s} + \b{g})\cdot \delta \b{R} \end{pmatrix} \b{r}_i = \omega_i \, \b{r}_i, \label{eq:firstOrderJacobi}
\end{align}
where
\begin{align}
h^{1,2}_\alpha &= \frac{\partial A_{12}^{I,I}}{\partial R_\alpha} \Big\vert_{\nucl_0}, \\
h^{2,1}_\alpha &= \frac{\partial A_{21}^{I,I}}{\partial R_\alpha} \Big\vert_{\nucl_0},
\end{align}
and
\begin{align}
s_\alpha &= \frac{1}{2} \Bigl( \frac{\partial A_{11}^{I,I}}{\partial R_\alpha} \Big\vert_{\nucl_0} + \frac{\partial A_{22}^{I,I}}{\partial R_\alpha} \Big\vert_{\nucl_0} \Bigr), \\
g_\alpha &= \frac{1}{2} \Bigl( \frac{\partial A_{22}^{I,I}}{\partial R_\alpha} \Big\vert_{\nucl_0} - \frac{\partial A_{11}^{I,I}}{\partial R_\alpha} \Big\vert_{\nucl_0} \Bigr).
\end{align}
The difference in energy of the states is
\begin{align}
\omega_2 - \omega_1 = 2 \sqrt{(\b{g}\cdot \delta \nucl)^2 + (\b{h}^{1,2}\cdot \delta \nucl)(\b{h}^{2,1}\cdot \delta \nucl)}.
\end{align}

In the nondefective case, $\J$ is similar to a symmetric matrix $\b{B}$. Let $\b{Q}$ be the matrix that relates $\J$ to $\b{B}$, that is, $\J = \b{Q}^{-1} \b{B} \, \b{Q}$. Since $\b{l}_1^T = \b{r}_1^T \b{Q}^T_0 \, \b{Q}_0$, the off-diagonal $A_{12}$ element can be written
\begin{align}
\begin{split}
A_{12}^{I,I} &= \b{l}_1^T \, \J \, \b{r}_2= \b{r}_1^T \b{Q}^T_0 \, \b{Q}_0 \, \b{Q}^{-1} \b{B} \, \b{Q} \, \b{r_2}. \label{eq:a12}
\end{split}
\end{align}
Above and in the following, we let $\b{Q}_0$ and $\b{B}_0$ denote the value of $\b{Q}$ and $\b{B}$ at $\nucl_0$, reserving $\b{Q}$ and $\b{B}$ for their value at $\nucl = \nucl_0 + \delta \nucl$. Let us expand $\b{Q}^{-1} \b{B} \, \b{Q}$ about $\nucl_0$:
\begin{align}
\begin{split}
\b{Q}^{-1} \b{B} \, \b{Q} &= \b{Q}_0^{-1} \b{B}_0 \, \b{Q}_0 \\
&\quad+ \frac{\partial}{\partial \b{R}} (\b{Q}^{-1} \b{B} \, \b{Q})\vert_{\nucl_0} \cdot \delta \nucl + \ldots
\end{split}
\end{align}
By defining $\J_0 = \b{Q}^{-1}_0 \b{B}_0 \b{Q}_0$, we can write
\begin{align}
\begin{split}
\frac{\partial}{\partial \b{R}} (\b{Q}^{-1} \b{B} \, \b{Q})\vert_{\nucl_0} &= \b{Q}_0^{-1} \frac{\partial \b{B}}{\partial \b{R}}\vert_{\nucl_0} \, \b{Q}_0 \\
&+ \frac{\partial \b{Q}^{-1}}{\partial \nucl}\vert_{\nucl_0} \b{Q}_0 \J_0 \b{Q}_0^{-1} \b{Q}_0 \\
&+ \b{Q}_0^{-1} \b{Q}_0 \J_0 \b{Q}_0^{-1}  \frac{\partial \b{Q}}{\partial \nucl}\vert_{\nucl_0} 
\end{split}
\end{align}
Inserting this expression for the derivative into Eq.~\eqref{eq:a12},
\begin{align}
\begin{split}
& A_{12}^{I,I} = \sum_\alpha \b{r}_1^T \b{Q}_0^T \frac{\partial \b{B}}{\partial R_\alpha}\vert_{\nucl_0} \, \b{Q}_0 \, \b{r}_2 \, \delta R_\alpha + \\
& \omega \sum_\alpha \b{l}_1^T \bigl(\frac{\partial \b{Q}^{-1}}{\partial R_\alpha}\vert_{\nucl_0} \b{Q}_0 + \b{Q}_0^{-1} \frac{\partial \b{Q}}{\partial R_\alpha}\vert_{\nucl_0} \bigr) \b{r}_2 \, \delta R_\alpha + \ldots
\end{split}
\end{align}
The second term vanishes by the product rule:
\begin{align}
\begin{split}
 A_{12}^{I,I} &= \sum_\alpha \b{r}_1^T \b{Q}_0^T \frac{\partial \b{B}}{\partial R_\alpha}\vert_{\nucl_0} \, \b{Q}_0 \, \b{r}_2 \, \delta R_\alpha \\
& + \omega \sum_\alpha \b{l}_1^T \frac{\partial}{\partial R_\alpha} (\b{Q}^{-1} \b{Q})\vert_{\nucl_0} \, \b{r}_2 \, \delta R_\alpha + \ldots \\
&= \sum_\alpha \b{r}_1^T \b{Q}_0^T \frac{\partial \b{B}}{\partial R_\alpha}\vert_{\nucl_0} \, \b{Q}_0 \, \b{r}_2 \, \delta R_\alpha + \ldots
\end{split}
\end{align}
We have thus found that 
\begin{align}
h^{1,2}_\alpha = \b{r}_1^T \b{Q}_0^T \frac{\partial \b{B}}{\partial R_\alpha}\vert_{\nucl_0} \, \b{Q}_0 \, \b{r}_2.
\end{align}
An analogous derivation shows that
\begin{align}
h^{2,1}_\alpha = \b{r}_2^T \b{Q}_0^T \frac{\partial \b{B}}{\partial R_\alpha}\vert_{\nucl_0} \, \b{Q}_0 \, \b{r}_1,
\end{align}
showing that ${\b{h}^{1,2} = \b{h}^{2,1}}$ by the symmetry of the matrix $\b{Q}_0^T (\partial \b{B})(\partial R_\alpha)\vert_{\nucl_0} \, \b{Q}_0$. It follows that
\begin{align}
\omega_2 - \omega_1 = 2 \sqrt{(\b{g}\cdot \delta \nucl)^2 + (\b{h}^{1,2}\cdot \delta \nucl)^2},
\end{align}
which is the well-known linearity of the energy gap obtained in symmetric theories\citep{Zhu2016} and more generally in exact quantum theory.\citep{Teller1937} We therefore expect that nonsymmetric theories that are nondefective have the correct energy gap linearity in the branching space close to the conical intersection.

\section{The description of accidental same-symmetry intersections} \label{sec:nosym}
At accidental same-symmetry conical intersections, also known as no-symmetry conical intersections, neither of the crossing conditions are satisfied by group theoretical arguments.\citep{Zhu2016} Coupled cluster theory has been found to be defective at this class of intersections.\citep{Kohn2007}  As discussed by Hättig, a degeneracy is then obtained in a subspace of dimension $N-1$ where\citep{Hattig2005} 
\begin{align}
\omega_2 - \omega_1 = \sqrt{(J_{22} - J_{11})^2 + 4 \, J_{12}\,J_{21}} = 0. \label{eq:defectCondition}
\end{align}

\begin{figure}[htb]
\includegraphics[width=\linewidth]{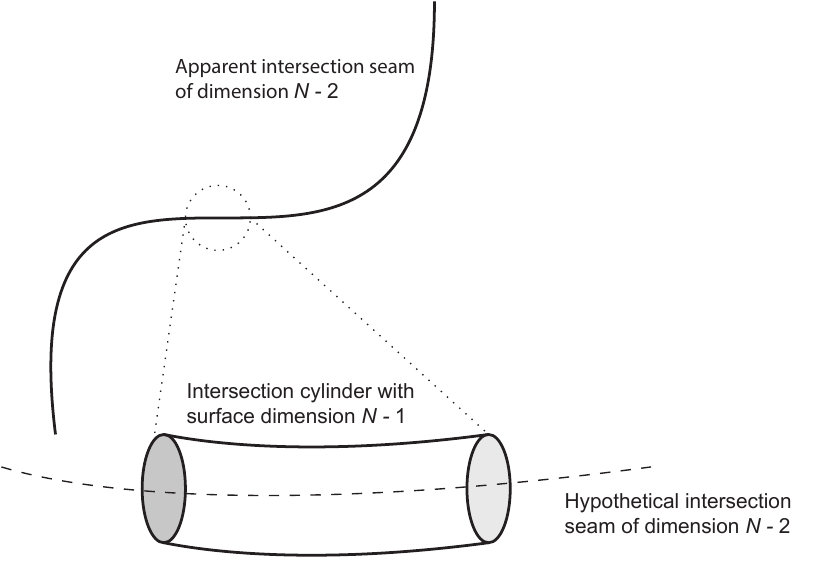}
\caption{\label{fig:cc_intersections} Accidental same-symmetry conical intersections using coupled cluster theory. Illustrated objects are for three vibrational degrees of freedom ($N = 3$).}
\end{figure}

The intersection defined by Eq.~\eqref{eq:defectCondition} is folded such that it resembles, from the perspective of large changes in $\nucl$, an object of dimension $N-2$. An illustration for $N = 3$ is given in Figure \ref{fig:cc_intersections}.
The reason for this is that the eigenvalues of $\J$ will converge to the eigenvalues of $H-E_0$ as more excitations are included in the cluster operator.\citep{Helgaker2014} For a given intersection of $H$, the complex energies predicted by $\J$ must eventually vanish, leaving the intersection of $H$ in the limit of a complete $T$. Consider the case $N = 3$. An intersection of $H$, for this number of internal degrees of freedom, is a curve in $\nucl$ space ($N-2 = 1$). The coupled cluster intersection, on the other hand, while resembling a curve from the perspective of large changes in $\nucl$, is in reality a cylinder whose surface has the dimensionality expected in light of Eq.~\eqref{eq:defectCondition} ($N-1 = 2$). On its surface the eigenvectors are parallel, and in its interior the energies are complex. The cylinder shrinks to a curve of dimension $N-2$ as all excitations are included in $T$ (\emph{i.e.}, to the conical intersection predicted by $H$).

For completeness, we note that there may exist a space of dimension $N-3$ where
\begin{align}
J_{11} = J_{22}, \quad J_{12} = 0, \quad J_{21} = 0.
\end{align}
For $N = 3$, this space corresponds to places along the intersection seam where the cylinder shrinks to a point. The subspace is of interest because some authors have identified it as the intersection.\citep{Hattig2005,Kohn2007} This is correct if $\J$ is nondefective, but in that case the number of independent conditions reduces to two (see Section \ref{sec:dimensionality}). When $\J$ is defective, the seam should instead be identified with the ($N-1$)-dimensional space, as this is consistent with the behavior of the model in the complete $T$ limit.

\section{The description of accidental symmetry allowed intersections} \label{sec:symallowed}

When the off-diagonal conditions are satisfied by group theoretical arguments, the intersection is known as accidental symmetry allowed.\citep{Zhu2016} An example is the $1 \; A{''}/2\; A{''}$ intersection in \ce{SH2} ($C_s$), which is located in the subspace of geometries where the molecule has $C_{2v}$ symmetry.\citep{Yarkony1996} For such $\b{R}$, the states possess $B_1$ and $A_2$ symmetry, and the off-diagonal elements of the totally symmetric $H$ vanish due to symmetry. This is also true in coupled cluster theory, where $\b{l}^T \J \, \b{r} = 0$ for $\b{l}$ and $\b{r}$ of different symmetry.

At symmetry allowed intersections, $\b{r}_1$ and $\b{r}_2$ do not become parallel since they possess different symmetries. It follows that $\b{J}$ is diagonal, and therefore nondefective, at the intersection. Expanding the matrix about a point of intersection, $\nucl = \nucl_0 + \delta \nucl$, where the displacement $\delta \nucl$ preserves the molecule's $C_{2v}$ symmetry, $\b{l}_1^T \J \, \b{r}_2 = 0$. Only the diagonal condition needs to be met (by accident, as far as symmetry is concerned) in the subspace $\mathcal{D} \subset \real^{N}$ of $C_{2v}$ geometries. There is thus one condition in $\mathcal{D}$, the same as in the symmetric case.\citep{Yarkony2001b}

Partitioning the displacement $\delta \nucl$ in a totally symmetric part, $\delta \nucl_{s}$, and a non-totally symmetric part, $\delta \nucl_{n}$, the eigenvalues can be written to first order as follows:
\begin{align}
\Delta \omega = \frac{1}{2}\sqrt{(\b{g}_s \cdot \delta \nucl_{s})^2 + (\b{h}_n^{1,2} \cdot \delta \nucl_n)(\b{h}_n^{2,1}\cdot\delta \nucl_n)}.
\end{align}
To derive the above, we made use of $\b{g}_n = 0$, which follows because $\partial \J / \partial \nucl_n$ does not span $A_1$, and ${\b{h}^{1,2}_s = \b{h}^{2,1}_s = 0}$. The latter is seen by noting that the two states possess different symmetry for totally symmetric displacements. 

In the \ce{SH2} system, there is associated with $\b{g}$ and $\b{h}$ one totally symmetric ($\delta R_s$) and one non-totally symmetric displacement ($\delta R_n$), respectively, giving
\begin{align}
\Delta \omega = \frac{1}{2}\sqrt{g_s^2 \, \delta R_{s}^2 + h_n^{1,2} \, h_n^{2,1} \delta R_n^2},
\end{align}
where $\b{h}^{1,2}$ and $\b{h}^{2,1}$ are parallel due to their orthogonality to both $\b{s}$ and $\b{g}$. An imaginary pair of excitation energies can result if $h_n^{1,2} \, h_n^{2,1}$ becomes negative. Although this is possible in principle, we have not found it to occur in practice (for \ce{SH2}), as shown below in Section \ref{sec:num_examples}.

In general, we may encounter defects when $\delta \nucl$ breaks the symmetry of the point group ($\nucl \notin \mathcal{D}$) and the states start to span the same symmetry. Loss of similarity to a symmetric matrix means that the energy gap linearity in the branching plane, as well as the uniqueness of $\b{h}$, may be lost for this class of intersections (see Section \ref{sec:perturbation}).

\section{Computational details}
All calculations were carried out using the Dalton quantum chemistry program.\citep{DALTON} The CCSD\citep{Purvis1982,Koch1990,Stanton1993a} energies were obtained with Dunning's augmented correlation consistent double-$\zeta$ basis (aug-cc-pVDZ).\citep{Dunning1989} 

For the excited states of hypofluorous acid (\ce{HOF}) and hydrogen sulfide (\ce{SH2}), the residuals were converged to within $10^{-5}$. For HOF, we performed the scan 
\begin{align}
\begin{split}
R_{\mathrm{OH}} &= 1.0900:0.0005:1.0950 \; \text{Å}, \\
R_{\mathrm{OF}} &= 1.3050:0.0002:1.3084 \; \text{Å}, \\
\theta_{\mathrm{HOF}} &= 90.50:0.05:91.50^\circ.
\end{split}
\end{align}
For \ce{SH2}, the investigated intersection point is
\begin{align}
\begin{split}
R_{\mathrm{SH}} &= 1.5092 \; \text{Å}, \\
\theta_{\mathrm{HSH}} &= 93.7689^\circ,
\end{split}
\end{align}
and the scan we performed in $\b{g}$ and $\b{h}$ as follows:
\begin{align}
\begin{split}
g &= -0.2500:0.0125:0.2500, \\ 
h &= -0.0500:0.0025:0.0500.
\end{split}
\end{align}
The vectors $\b{h}$ and $\b{g}$ are given in Table \ref{tab:gh}. The plots in Figures \ref{fig:symmetryallowed}, \ref{fig:realPartHOF}, and \ref{fig:imagPartHOF} are of interpolated values.

\begin{table}
\caption{Coordinates of $\b{g}$ and $\b{h}$ at the intersection point of \ce{SH2} (see text). The molecule lies in the $xz$-plane with \ce{S} positioned at the origin and \ce{H1} on the $z$-axis.}
\begin{ruledtabular}
\begin{tabular}{lccc}
Atom & $q$ & $g_q$ & $h_q$ \\
\hline
S & $x$ & 0.00000000 & 0.00000000 \\
  & $y$ & 0.00000000 & 0.00000000 \\
  & $z$ & 0.00000000 & 0.00000000 \\
 \hline
\ce{H1} & $x$ & -0.00757588 & 0.00000000 \\
		& $y$ & 0.00000000 & 0.00000000 \\
		& $z$ & -0.11145709 & 0.25668976 \\
\hline
\ce{H2} & $x$ & -0.11171402 & -0.25613461 \\
		& $y$ & 0.00000000 & 0.00000000 \\
		& $z$ & -0.00023309 & 0.01687298 \\
\end{tabular}
\end{ruledtabular} \label{tab:gh}
\end{table}

\section{Numerical examples} \label{sec:num_examples}
We study two systems with three vibrational degrees of freedom, \ce{HOF} and \ce{SH2}. These molecules provide simple illustratations of the arguments in Sections \ref{sec:crossing}--\ref{sec:symallowed}, as the intersection is a curve for both systems (see Figure \ref{fig:cc_intersections}).

\subsection{The $2\,{^{1}}A{'} / 3\,{^{2}}A{'}$ accidental same-symmetry conical intersection of hypofluorous acid (HOF)}
An intersection between the first two singlet excited states of $A'$ symmetry in hypochlorous acid (\ce{HClO}) was identified by Nanbu and Ivata some two decades ago.\citep{Nanbu1992} Here we study the analogous intersection between the 2$^1$A$'$ and 3$^1$A$'$ states of hypofluorous acid (\ce{HOF}). 

In Figure \ref{fig:realPartHOF} we show a series of slices of $\nucl$ space, in the $R_\mathrm{OH}$ direction, where the coloring corresponds to the real part of the energy difference between the two states. This difference becomes zero at the seam, which has the shape of a filled cylinder, as shown in Figure \ref{fig:cc_intersections} and discussed in Section \ref{sec:nosym}. 

The real part of the energy difference vanishes at its surface, and an imaginary pair is created in its interior. In Figure \ref{fig:imagPartHOF}, the magnitude of the imaginary part of the complex pair is shown for the same slices in $R_\mathrm{OH}$. The extent of the cylinder is approximately $0.5^\circ$ in the \ce{H-O-F} angle and $0.0010 \; \text{Å}$ in the $R_\mathrm{OF}$ direction.

\subsection{The $1 \; {^{1}}A{''}/2\; {^{1}}A{''}$ accidental symmetry allowed intersection in hydrogen sulfide (\ce{SH2})}
The $1 \; {^{1}}A{''}/2\; {^{1}} A{''}$ symmetry allowed intersection of \ce{SH2} is a standard example of this class of intersections.\citep{Yarkony2001b}

The vectors $\b{g}$ and $\b{h}$ were determined as follows. Since both $\b{g}$ and $\b{s}$ only have components in the two $A_1$ modes, a search in this plane provided $\b{s}$, as the direction that preserved the degeneracy, and $\b{g}$, as the direction orthogonal to $\b{s}$ in the $A_1$ space; $\b{h}$ was then determined as the vector orthogonal to both $\b{s}$ and $\b{g}$.

In Figure \ref{fig:symmetryallowed}, the branching plane is shown at a point of intersection along the seam. We did not observe complex energies, indicating that $\J$ is nondefective in this region of $\nucl$. Linearity to first order in $\b{g}$ and $\b{h}$ appears valid, in agreement with the analysis given in Section \ref{sec:perturbation}.

\section{Concluding remarks}
By reconsidering the eigenvalue problem in coupled cluster theory, we have shown that the correct number of crossing conditions are predicted so long as the coupled cluster Jacobian is nondefective. With this property, the theory is expected to give a proper description of conical intersections, with the correct conical shape of the energy surfaces to first order in the branching plane.

However, the Jacobian matrix is defective at accidental same-symmetry conical intersections. As we have demonstrated in calculations on hypofluorous acid, the observed defective intersection seam is a higher-dimensional surface ($N-1$) folded about an ($N-2$)-dimensional space. In the limit of a complete cluster operator, the dimension reduces to $N-2$, indicating that minor modifications are needed to allow a correct description of intersections.

In a recent paper, we were indeed able to remove defects in the Jacobian matrix by appropriately modifying the coupled cluster model.\citep{Kjoenstad2016} 

Symmetry ensures that the Jacobian is nondefective at accidental symmetry allowed intersections, though not necessarily in their vicinity. At a conical intersection of this class in hydrogen sulfide, we found coupled cluster theory to be nondefective and have the correct first order linearity of the energy gap in the branching plane. 

\begin{acknowledgments}
We thank Robert M. Parrish and Xiaolei Zhu for enlightening discussions in the early stages of the project.
Computer resources from NOTUR project nn2962k are acknowledged. Henrik Koch acknowledges financial support from the FP7-PEOPLE-2013-IOF funding scheme (Project No. 625321). Partial support for this work was provided by the AMOS program within the Chemical Sciences, Geosciences, and Biosciences Division of the Office of Basic Energy Sciences, Office of Science, US Department of Energy. We further acknowledge support from the Norwegian Research Council through FRINATEK project no. 263110/F20.
\end{acknowledgments}

\bibliographystyle{aip}
\bibliography{library}

\end{document}